# A framework for linking literature-based knowledge integration and infrastructure-supported knowledge integration: Opportunities and challenges from a case study

Mahlet Degefu Awoke*[1], Hadi Ghaemi[2], Lauren Snyder[2], Markus Stocker[2], Tilman Brück[1,3,4]

[1]Leibniz Institute of Vegetable and Ornamental Crops (IGZ), Großbeeren, Germany

[2]TIB - Leibniz Information Centre for Science and Technology, Hanover, Germany

[3]Humboldt-Universität zu Berlin, Berlin, Germany

[4]International Security and Development Center (ISDC), Berlin, Germany

*Corresponding author: awoke@igzev.de

## Abstract

Integrating knowledge across disciplines is central to sustainability research, yet most evidence-synthesis methods rely on findings as reported in publications, limiting verification and reuse of underlying data and workflows. We develop a conceptual framework linking literature-based and infrastructure-supported knowledge integration, using a systematic review case study to examine when integration can extend beyond reported findings. We reviewed 37 studies on climate change, violent conflict, and household food security. Literature-based synthesis enabled integration across all included studies, whereas access to reusable outputs was limited: over half provided no data availability statement, 27% reported availability upon request, but reusable data and workflows were available for only 8%. To explore infrastructure-supported integration, we used the TIB Knowledge Loom to represent studies with accessible data and code as machine-readable outputs, and produced a knowledge gap map (KGM) from manually extracted and Loom-derived data, comparing manual and infrastructure-supported synthesis. Where outputs were reusable, synthesis could be produced directly from data and workflows rather than from publications. These findings show that literature-based synthesis can be complemented by infrastructure-supported integration where outputs are accessible and usable, and that advancing knowledge integration depends not only on infrastructures but on making data, code, and workflows accessible, executable, and reusable.

**Keywords**: Data reusability; Evidence synthesis; Knowledge gap map; Open science; Research data infrastructure; Systematic reviews

## 1 Introduction

Addressing sustainability challenges requires integrating knowledge across different disciplines and analytical approaches. Climate, biodiversity, agriculture, and food constitute a closely interconnected area of action and conflict: changes in one domain can generate consequences, trade-offs, and feedbacks across the others. These interactions are shaped by environmental, economic, and political processes and may contribute to mutually reinforcing crises, such as polycrises - that cannot be adequately understood through any single disciplinary perspective (Brandt et al., 2013). Addressing such polycrises, therefore, requires approaches that combine evidence from multiple processes and disciplines to analyse these interconnected challenges (Lang et al., 2012; Pohl et al., 2021). Knowledge integration is central to this effort, as it links theoretical perspectives, empirical evidence, and methodological approaches to generate more comprehensive analyses of complex socio-ecological systems and polycrises (Hoffmann et al., 2017).

A standard mechanism for knowledge integration is literature-based integration. Methods such as systematic reviews, meta-analyses, and evidence maps provide structured procedures for identifying relevant studies, applying transparent inclusion criteria, and comparing findings across a heterogeneous body of literature, even within single disciplines (Page et al., 2022; Yu et al., 2022). These approaches are central to knowledge integration because they standardize workflows, supporting transparency and comparability across studies. At the same time, these approaches typically rely on findings reported in scientific publications, meaning that knowledge integration is based primarily on textual descriptions of methods, variables, and results with limited direct access to the data and analytical processes that produced them (Depraetere et al., 2021; Ghaemi et al., 2025).

However, integrating knowledge across studies faces several recurring challenges. First, the rapid growth and fragmentation of scientific literature make comprehensive coverage increasingly difficult (Rethlefsen et al., 2024). Second, methodological heterogeneity across studies, including differences in indicators, data sources, and analytical approaches, complicates systematic comparison and integration (Bornmann & Mutz, 2015; Cooper et al., 2019). This concern extends to the conduct of evidence synthesis itself: a recent scoping review of published mapping reviews found significant variability in their methodology and reporting, highlighting persistent gaps in evidence synthesis methodologies more broadly (Khalil et al., 2024). In addition, reliance on reported results makes knowledge integration dependent on authors' reporting choices, such as potential bias toward statistically significant results, underreporting of null findings, and variation in how results and significances are defined, interpreted, and presented (Dwan et al., 2013). More fundamentally, the research artefacts used to produce scientific findings, including data, analytical code, and computational workflows, are often insufficiently documented, limiting both verifiability of reported results and reuse (Ivimey-Cook et al., 2025; Pollack et al., 2026; Tedersoo et al., 2021).

Recent developments in open science and digital research infrastructures create the possibility of extending knowledge integration beyond the findings reported in publication texts by connected research outputs. The FAIR principles emphasize that data and related digital objects should be findable, accessible, interoperable, and reusable in order to support transparency and reproducibility (Wilkinson et al., 2016). Building on these principles, emerging infrastructures seek to connect publications with the datasets, code, and workflows on which scientific findings are based, and to represent these outputs in structured, machine-readable forms (Lezhnina et al., 2025; Stocker et al., 2025) . Where such outputs are accessible and reusable, synthesis may be able to proceed not only by interpreting reported findings but also by engaging directly with the analytical artefacts from which they were derived.

Despite these developments, we still know relatively little about how literature-based knowledge integration and infrastructure-supported approaches can be combined in practice. While systematic reviews make it possible to integrate evidence across a broad body of literature, infrastructure-supported approaches depend on whether datasets and analytical workflows are actually available and usable (Huber et al., 2021). This leaves open the question to what extent these two forms of integration complement one another, and under what conditions infrastructure-supported approaches can move knowledge integration beyond what is reported in publications.

Given this knowledge gap, we develop and apply a conceptual framework linking literature-based integration with infrastructure-supported approaches. We distinguish between two complementary modes of integration: (i) literature-based integration, which draws on reported results in scientific publications, and (ii) infrastructure-supported integration, which enables verification and reuse when underlying datasets and analytical workflows are accessible (Stocker et al., 2025). Using a systematic review as an entry point, we examine how far integration can move beyond publication-level evidence and where its practical limits remain.

We illustrate our framework with a case study based on a systematic literature review of relationships between climate change, violent conflict, and household food security. While the substantive, topical findings of that review are reported elsewhere (Awoke & Brück, 2026a), our focus here is methodological, outlining the knowledge gaps in knowledge integration and lessons learnt from our framework and our case study. By reflecting on our case study with a bird's-eye perspective, we use the review process to examine when and how infrastructure-supported integration can complement conventional synthesis, and where it remains constrained by the availability of reusable research outputs. In doing so, we highlight both the opportunities and the limits of extending knowledge integration beyond published findings.

We find that literature-based integration was possible across the full review corpus, whereas infrastructure-supported integration was feasible only for a small subset of studies because reusable data and analytical workflows were rarely accessible. Where such outputs were available, however, they made it possible to produce synthesis outputs directly from data and workflows rather than reconstructing them from published findings. Taken together, these observations point to both the potential of infrastructure-supported integration and the practical limits imposed by current sharing and documentation practices.

Our paper contributes to the literature in three ways. First, it contributes to evidence synthesis research by clarifying the scope and limits of literature-based approaches for interdisciplinary knowledge integration. Second, it contributes to scholarship on open science and digital infrastructures by showing under what conditions reusable data, code, and workflows can extend knowledge integration beyond reported findings. Third, it contributes to sustainability research methodology by developing a conceptual framework for understanding how literature-based approaches and infrastructure-supported knowledge integration relate to and can complement each other under different practical conditions.

# 2 Conceptual framework for knowledge integration

## 2.1 Data and knowledge

Understanding knowledge integration requires conceptual clarity about the elements that constitute scientific knowledge and how they relate to one another. Data refer to empirical observations or measurements collected through research, often recorded as structured entries in databases or digital repositories (Leonelli, 2019), whereas knowledge emerges from the interpretation and contextualization of such data within theoretical and methodological frameworks (Wolski & Gomolińska, 2020). In this sense, knowledge represents structured understanding derived from empirical evidence and analytical reasoning. Floridi sharpened the distinction by defining semantic information as well-formed, meaningful, and truthful data, and knowledge as accounted, that is, justified, semantic information. Scientific knowledge is systematically accounted semantic information, since the accounting process follows the scientific method. Scientific articles contain scientific knowledge because the semantic information reported in articles is accounted for and the accounting process is systematic (Floridi, 2011). The distinction between data and knowledge is particularly relevant for understanding how research outputs can be reused, as access to underlying data and analytical processes determines whether knowledge claims can be verified and integrated across studies.

We illustrate the distinction with a simple example. Assume the Iris dataset with (measurement) data about petal and sepal length of three Iris species. The statement that the petal lengths of Setosa and Virginica are significantly different constitutes semantic information, as the linguistic statement is well-formed, meaningful, and truthful data. A statistician can provide an account, in terms of a data analysis, for this specific semantic information, e.g., in the form of an R script that implements the statistical test and reports the p-value. Thus, the statistician knows that Setosa and Virginica petal lengths are statistically significantly different from each other because she can account for the piece of semantic information (Fisher, 1936; Unwin & Kleinman, 2021). The statistician may report this piece of semantic information together with the account in her next research article, which thus contains knowledge. This example illustrates how scientific knowledge can be represented not only in published findings but also in the analytical account through which those findings are produced. This matters for the present paper because knowledge integration may draw either on reported claims in publications or, where available, on the data and analytical workflows that account for those claims.

### 2.2 Knowledge integration

Knowledge integration has been conceptualized in different ways, reflecting the different purposes it may fulfil across research contexts. In this paper, we use the term to describe knowledge processes through which different forms of knowledge are connected in order to generate new understanding of complex problems and to support the development of responses to real-world challenges by identifying knowledge gaps and connecting findings across theory, empirical data, and practice (Hoffmann et al., 2017; Lang et al., 2012; Pohl et al., 2021). In terms of Floridi's definition, knowledge integration can be understood as the integration of systematically accounted semantic information. The integration of semantic information reported in research articles is thus an example of (scientific) knowledge integration (Floridi, 2011). In sustainability research, knowledge integration is particularly important because many societal challenges involve interactions among environmental, social, and economic systems and extend across scientific disciplines and societal sectors. Understanding and addressing these challenges therefore requires not only the integration of knowledge across disciplines but also transdisciplinary approaches that connect scientific knowledge with the perspectives and experiential knowledge of actors from policy, civil society, and practice (Lang et al., 2012; Norström et al., 2020). The scientific knowledge integration examined in this paper represents one component of this broader landscape.

Importantly, integration is not a purely additive process in which information from different sources is simply aggregated. Rather, it is often described as an iterative learning process in which knowledge components are interpreted, related, and sometimes reconfigured to generate new insights (Karrasch et al., 2022; Rousseau et al., 2019). Through this process, different forms of knowledge, including disciplinary expertise, empirical evidence, and experiential knowledge, are combined to support a more comprehensive understanding of complex systems.

Knowledge integration in sustainability research often takes place in interdisciplinary and transdisciplinary settings. Interdisciplinary research combines theoretical and methodological perspectives from multiple academic disciplines in order to address complex research questions (Tobi & Kampen, 2018), while transdisciplinary research extends beyond disciplinary boundaries by incorporating knowledge from societal actors, practitioners, and stakeholders into the research process (Hadorn et al., 2008; Lang et al., 2012). However, knowledge integration is considered a major challenge of inter- and transdisciplinary research (Hoffmann et al., 2017). Both forms are particularly relevant in sustainability research, where addressing complex societal challenges often requires combining scientific expertise with experiential and practice-based knowledge (Hadorn et al., 2008; Pohl et al., 2021; Tobi & Kampen, 2018). One important dimension of this challenge concerns how knowledge is represented, documented, and made accessible across studies and domains. When knowledge remains embedded in disciplinary formats or is insufficiently documented, it becomes difficult to connect findings, compare results, or reuse analytical approaches across contexts.

In this paper, we focus on this dimension of knowledge integration: how research findings are represented and made accessible across studies. This shifts attention from integration as a purely conceptual or collaborative process to integration as a function of how research outputs are organized, accessed, and can be reused. While knowledge integration has often been discussed as an interactive and open-ended process in inter- and transdisciplinary research (Pohl et al., 2021), it has also been identified as a central challenge in transdisciplinary synthesis processes (Hoffmann et al., 2017). A related approach is Open Synthesis, which refers to the application of Open Science principles including open methods, open data, and open access throughout the evidence-synthesis process (Haddaway, 2018). It calls for knowledge integration to embrace these principles to improve the reliability, trustworthiness, and reuse of synthesized information (Haddaway, 2018). Against this background, we distinguish two approaches to knowledge integration across disciplines: literature-based integration of reported findings and infrastructure-supported integration based on reusable research outputs.

### 2.3 A Framework for knowledge integration: Literature-based and infrastructure-supported integration

Building on the conceptual discussion above, we develop a framework for knowledge integration based on recurring challenges identified (Awoke & Brück, 2026a; Barker et al., 2022; Laurinavichyute et al., 2022; Rethlefsen et al., 2024; Wilkinson et al., 2016). Taken together, these challenges point to four key dimensions of knowledge integration across studies: coverage, structured synthesis, verifiability, and reusability. First, fragmentation and the rapid growth of the literature create a challenge of coverage because relevant evidence is dispersed across domains and sources. Second, methodological and conceptual heterogeneity creates a challenge of structured synthesis because findings must be compared and organized systematically across studies. Third, reliance on published reporting creates a challenge of verifiability because reported results are difficult to assess without access to the analytical procedures that produced them. Fourth, limited accessibility and usability of datasets, code, and workflows create a challenge of reusability because deeper forms of integration depend on whether underlying research outputs can be directly reused.

We use these four dimensions: coverage, structured synthesis, verifiability, and reusability, to compare two complementary modes of knowledge integration: literature-based integration and infrastructure-supported integration. Literature-based integration is applicable across heterogeneous studies and is particularly suited to achieving coverage and structured synthesis (Pati & Lorusso, 2018). Infrastructure-supported integration can strengthen verifiability and reusability by linking publications to underlying datasets, code, and analytical workflows, but only where such outputs are accessible and sufficiently documented (Haddaway, 2018; Stocker et al., 2025).

Rather than treating these approaches as separate, the framework conceptualizes literature-based synthesis and infrastructure-supported integration as complementary mechanisms of knowledge integration. Figure 1 illustrates the relationship between literature-based synthesis and infrastructure-supported integration within this framework.

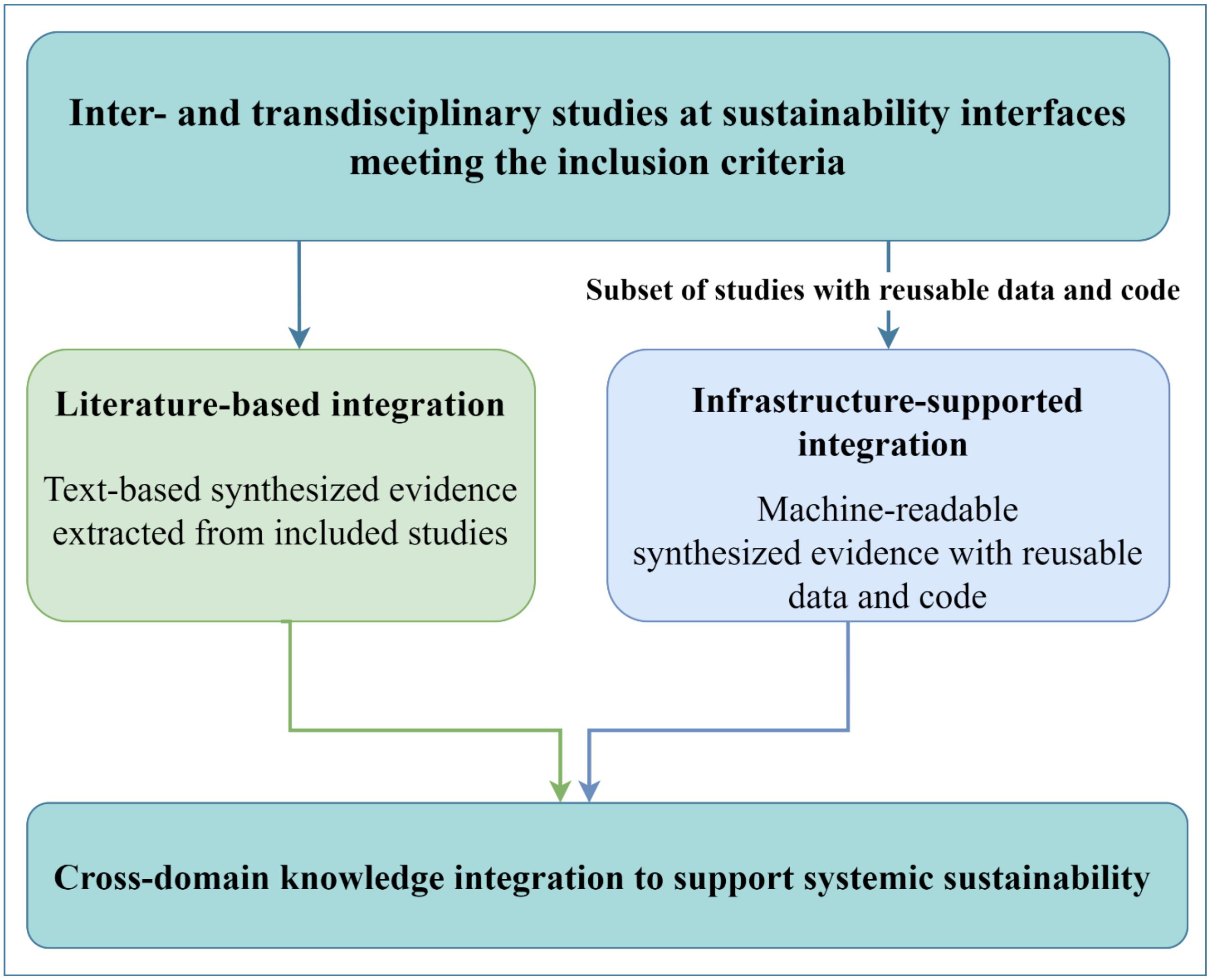

**Figure 1: Conceptual framework linking literature-based knowledge integration with infrastructure-supported knowledge integration.**

*Literature-based knowledge integration: Systematic literature review*

Methods such as systematic literature reviews provide structured procedures for identifying relevant studies, applying transparent inclusion criteria, and extracting comparable information across a body of literature (Pati & Lorusso, 2018). In this study, literature-based knowledge integration is implemented through a systematic literature review workflow that systematically identifies and organizes empirical evidence across studies. Because it relies on published articles, literature-based integration can be applied to the full set of studies that meet a review's inclusion criteria, regardless of whether the underlying datasets or analytical workflows are available.

This approach provides a basis for identifying patterns across studies, comparing empirical findings, and mapping research gaps. At the same time, integration remains based on reported results rather than on direct access to the analytical processes that underlie them. As a result, verification of analyses and reuse of computational workflows are generally not possible in this mode of integration. In other words, published articles do account for the reported information, but the account and the reported information are, in general, not reusable or easily reproducible for, or with support of, machines.

*Infrastructure-supported knowledge integration*

Infrastructure-supported integration aims to extend the text-based synthesis by incorporating research infrastructures that connect scientific publications with the underlying research outputs from which findings are derived. This form of integration becomes possible when studies provide reusable datasets, analytical code, and documentation that enable the reproduction of analytical procedures and their results. In this paper, this approach is illustrated using the TIB Knowledge Loom developed by the TIB - Leibniz Information Centre for Science and Technology, an emerging open science digital library of analysis-ready scientific knowledge (Stocker et al., 2025). However, the feasibility of infrastructure-supported integration, specifically the application of the TIB Knowledge Loom, depends on the availability and usability of research outputs.

*Linking the two approaches*

Within our framework, we position literature-based synthesis as the entry point for knowledge integration, as it enables us to identify and organize relevant studies and their reported findings. We then build on this foundation through infrastructure-supported integration, which allows us to access the research artefacts underlying (a subset of) these studies. Systematic literature review workflows, therefore, play a dual role.

In addition to supporting the synthesis of reported findings, it enables the identification of studies that provide reusable datasets and analytical workflows. In this sense, literature-based synthesis and infrastructure-supported integration are best understood as complementary mechanisms: the former provides broad coverage across studies, while the latter extends the depth of integration where reusable research artefacts are available.

# 3 Case study and methodological approach

### 3.1 Case study context

To illustrate the practical application of the proposed framework, we draw on a systematic literature review examining the relationships among climate change, violent conflict, and household food security. Research at this interface spans multiple disciplinary domains, including climate science, development studies, economics, and political science, and therefore represents a context in which integrating heterogeneous forms of evidence is particularly challenging. Such interfaces are also central to inter- and transdisciplinary sustainability research, where knowledge must often be connected across domains such as climate, agriculture, and food systems.

In this case study, we focus on quantitative survey-based studies that examine household-level food security outcomes in relation to climatic and conflict-related stressors. While we report the substantive findings of this systematic review elsewhere (Awoke & Brück, 2026a), the purpose of the case study here is different. Rather than revisiting the substantive results, we use the review workflow to examine how classical literature-based synthesis can be complemented by novel infrastructure-supported knowledge integration. The case study thus demonstrates the practical application of the proposed knowledge integration framework.

## 3.2 Systematic review workflow

*Search engines for literature identification*

We conducted literature searches using three major bibliographic databases, Scopus, Web of Science, and PubMed, which provide structured access to scientific literature across multiple disciplines (Gusenbauer & Haddaway, 2020). We constructed the search strings (Appendix 1) by combining terms related to climate change, climate shocks, violent conflict, and household food security to identify relevant empirical studies.

*Reference management tools*

We organized the retrieved records using the reference management software EndNote and Zotero, which facilitated the storage and organization of bibliographic records (Hammer et al., 2023). These tools support the structured documentation of search results and contribute to transparent and reproducible review workflows. For instance, we used Zotero to enable shared access to references within an open-source environment (Lorenzetti & Ghali, 2013).

*Screening and study selection*

We conducted the study-selection process following the PRISMA guidelines (Page et al., 2021), including duplicate removal, title and abstract screening, and full-text assessment. We used Rayyan to support duplicate removal and screening management. Rayyan is a web-based platform that facilitates systematic review workflows by enabling efficient screening and management of references (Hammer et al., 2023; Johnson & Phillips, 2018). We then applied ASReview, an open-source active learning system, to prioritize potentially relevant records during title and abstract screening (Khalil et al., 2024; Van De Schoot et al., 2021).

ASReview applies machine-learning algorithms to assist reviewers in identifying relevant studies while maintaining transparent documentation of screening decisions (Figure 2). Following the screening and full-text assessment stages, a final sample of 37 peer-reviewed studies met the inclusion criteria (Appendix 2). These studies constituted the evidence base for the literature-based knowledge integration and were subsequently assessed for infrastructure-supported knowledge integration.

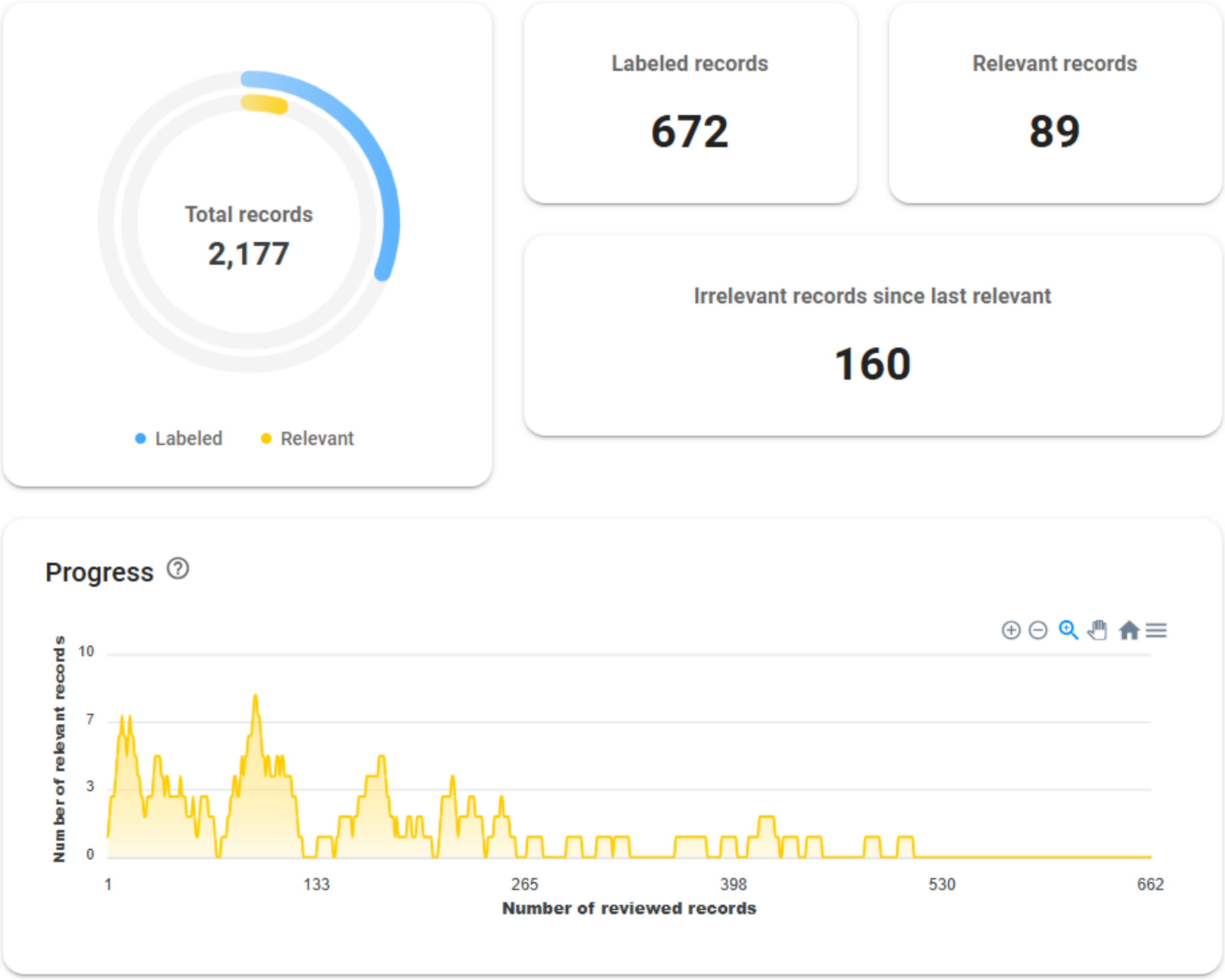

**Figure 2: Methodological output of the literature-based knowledge-integration workflow. ASReview screening progress, showing the number of labeled and relevant records and the prioritization of relevant studies during active learning–assisted screening.**

### 3.3 Assessing the feasibility of infrastructure-supported knowledge integration

The second component focuses on infrastructure-supported knowledge integration, which (in our setting) depends on access to underlying datasets and analytical code. To assess the feasibility of this mode of integration, we examined the availability of data and analytical workflows for all 37 studies included in the systematic review.

We assessed feasibility in three ways. First, we reviewed data availability statements reported in the published articles. Second, we examined publicly linked repositories to determine whether datasets or analytical code were accessible. Third, we contacted corresponding authors to request access to underlying datasets and analytical scripts where these were not publicly available.

### 3.4 Operationalizing infrastructure-supported integration

To explore the potential of infrastructure-supported knowledge integration, we applied the TIB Knowledge Loom in the case study. The platform enables linking scientific findings to their underlying datasets, analytical code, and procedures, thereby providing a structured representation of research outputs that supports transparency, reproducibility, and reuse (Stocker et al., 2025).

For studies that met the criteria for infrastructure-supported integration, we created Loom records linking scientific findings published originally in the related articles to the corresponding analytical workflow, underlying methods, data, and code. Prior to publishing these Loom records, we contacted the respective authors to review the entries and confirm their approval. We asked authors to review the Loom record and confirm licensing arrangements for the shared datasets and scripts.

As part of the literature-based synthesis workflow, we also generated a knowledge gap map (KGM) based on manually extracted and coded information from the included studies. A KGM is a structured evidence-mapping tool that organizes and visualizes how studies relate key variables and outcomes across a heterogeneous body of literature. Such maps are used to provide a systematic overview of the distribution of evidence across thematic domains, thereby allowing both concentrations of research and areas with limited or absent evidence to be identified (Snilstveit et al., 2017).

We systematically extracted variables such as the type of crisis, food security indicators, and the reported direction of relationships, and analysed them using an R-based workflow. This produced a manually constructed synthesis output based on information extracted from scientific publications. To further explore the potential of infrastructure-supported knowledge integration, we then applied the same analytical workflow to data obtained directly from relevant Loom records. Because the Loom records already contained structured data, we accessed the inputs required for the data analysis that produced the KGM without having to manually extract them from the original publication. This allowed us to compare a synthesis workflow based on manual extraction from publications with one based on structured and reusable research outputs.

# 4 Empirical findings

This section presents findings from the application of the proposed knowledge-integration framework. The substantive empirical findings of the literature-based synthesis are reported elsewhere (Awoke & Brück, 2026a). Here, we focus on the extent to which the 37 studies identified through that synthesis provided accessible and reusable data and analytical workflows, and on how these outputs enabled infrastructure-supported knowledge integration to complement the literature-based synthesis.

## 4.1 Data availability statements

As a first step, we reviewed the data availability statements reported in the included studies (Table 1). More than half of the studies (54%) did not include any data availability statement. Among those that did, the most common category was "data available upon request" (27%), while only a small share of studies reported that datasets and analytical code were publicly available (5%). An additional 5% of studies stated in their data availability statements that datasets were available, but did not provide accompanying analytical code, and 5% explicitly indicated that data were restricted or not available. These results provide an initial indication of the limited availability of reusable research outputs at the level of formal reporting. However, data availability statements do not necessarily reflect the actual accessibility or usability of research outputs in practice.

**Table 1: Data availability statement across studies (N=37)**

| **Data availability statement category** | **Count** | **Share (%)** |
|---|---|---|
| No data availability statement | 20 | 54 |
| Data available upon request | 10 | 27 |
| Data and code publicly available | 2 | 5 |
| Public dataset only (no code provided) | 2 | 5 |
| Data restricted / not available | 2 | 5 |
| Not applicable | 1 | 3 |
| **Total** | **37** | **100** |

### 4.2 Access to data and analytical workflows

To assess whether reported data availability translates into effective access, we contacted corresponding authors to request datasets and analytical code (Table 2). In total, we obtained reusable data and analytical workflows from only three studies (8%), either through public availability (5%) or upon request (3%). In contrast, most studies (81%) remained inaccessible due to non-response from authors.

**Table 2: Response rate of authors and outcome of data request**

| Category | Count | Share (%) | Author response | Outcome |
|---|---|---|---|---|
| Publicly available | 2 | 5 | Not required | Reusable |
| Provided and usable | 1 | 3 | Yes | Reusable |
| Provided but unusable | 1 | 3 | Yes | Not reusable |
| Responded, but data not accessible (lost or conditional) | 2 | 5 | Yes | Not accessible |
| Contact failure (email bounced) | 1 | 3 | Not reachable | Not accessible |
| No response | 30 | 81 | No | Not accessible |
| **Total** | **37** | **100** | | |

Even when authors responded, access was not straightforward. In one case, the data were no longer available; in another, access was conditional and therefore not granted. In addition, one study provided data and code that could not be executed successfully, and follow-up communication did not resolve the issue. In one further case, the contact email was no longer valid.

These results contrast with the data availability statements reported in Table 1. Although several studies indicated that data were available upon request, this rarely resulted in access to usable data in practice. This suggests that data availability statements do not reliably reflect the actual accessibility of reusable research artefacts. From the perspective of our framework, this implies that our proposed infrastructure-supported knowledge integration can be applied to only a very limited subset of studies.

### 4.3 Outputs of infrastructure-supported knowledge integration

Based on our assessment of data and code availability, we created three Loom records (Ceballos et al., 2026; George & Adelaja, 2026; Kafando & Sakurai, 2026) for studies that provided both datasets and executable analytical scripts. Figure 3 shows an example of such a record illustrating how scientific findings can be linked to the underlying analytical workflow, methods, data, and code through structured statements.

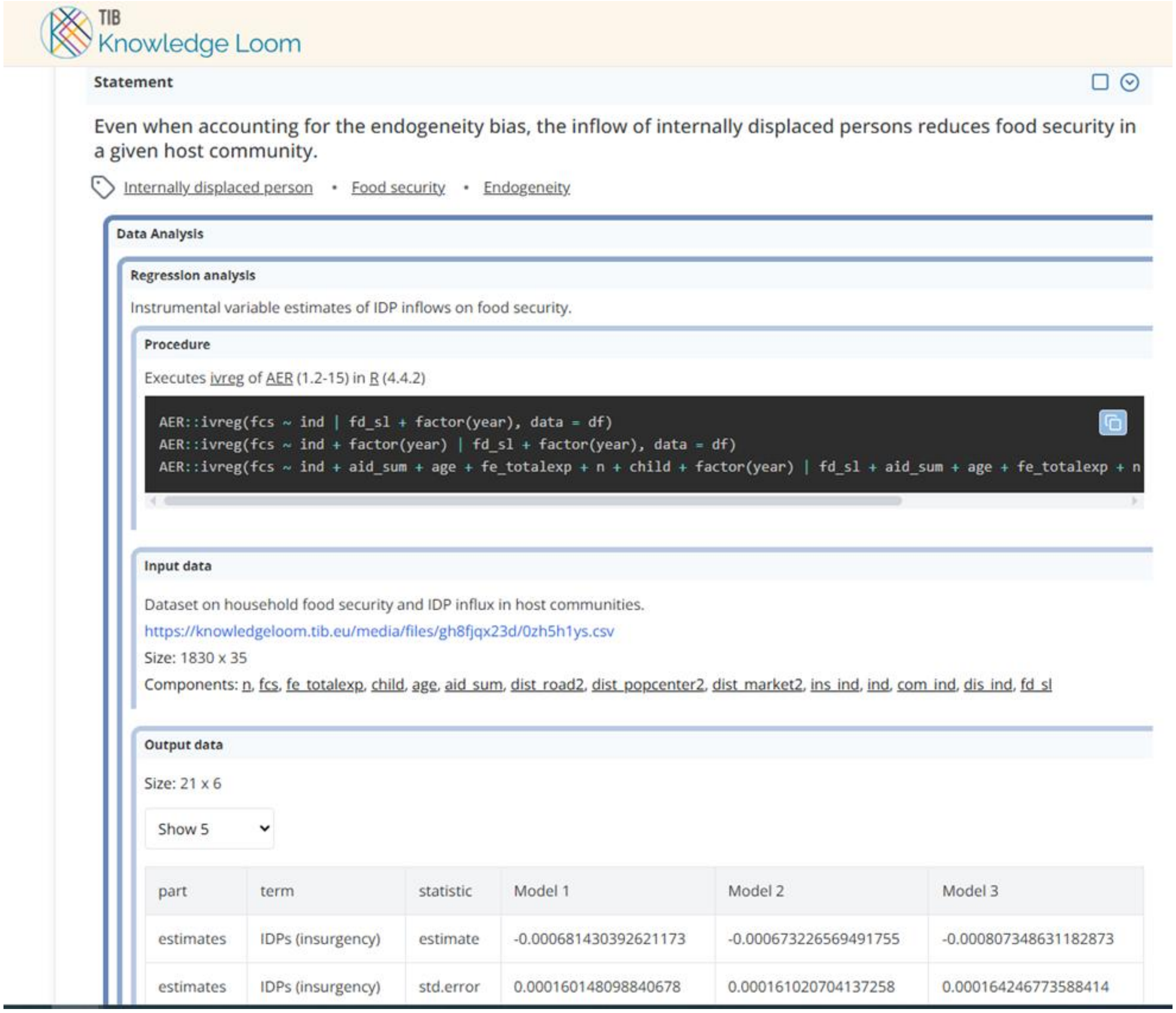

**Figure 3: Example of a Loom record linking scientific statements with underlying analytical workflows, methods, data, and code.**

As part of the classical systematic review workflow, we generated a knowledge gap map (KGM) based on manually extracted and coded information from the included studies (Figure 4). The resulting map provides a structured overview of the distribution of evidence and highlights areas where empirical evidence remains limited. In this study, the KGM also serves as an analytical interface for knowledge integration by structuring findings from diverse studies in a comparable format. However, its construction within a classical systematic review workflow depends on the manual extraction and coding of information from published articles.

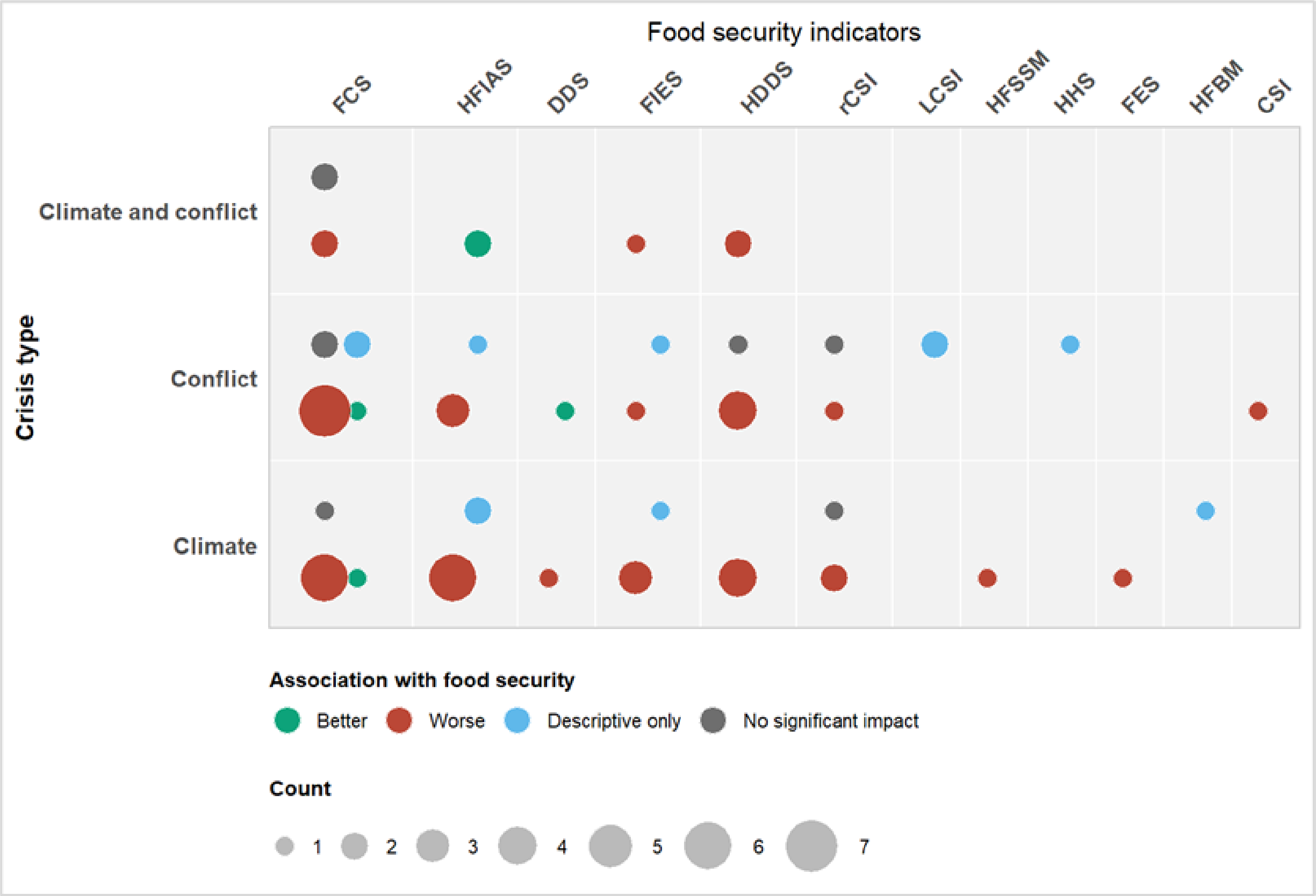

**Figure 4: Knowledge gap map based on manually extracted data from the systematic review (Awoke & Brück, 2026).**

When we applied the same analytical workflow to data obtained directly from a Loom record, the resulting output produced a partial knowledge gap map, an example of mini synthesis based on Loom-derived data (Figure 5). Although demonstrated using only three studies, this example illustrates how synthesis outputs, such as knowledge gap maps, can be generated directly from structured, reusable research outputs. In contrast to the manual workflow, which depends on extracting and interpreting information from publications, this approach enables the reuse of data and analytical procedures as provided in the original study.

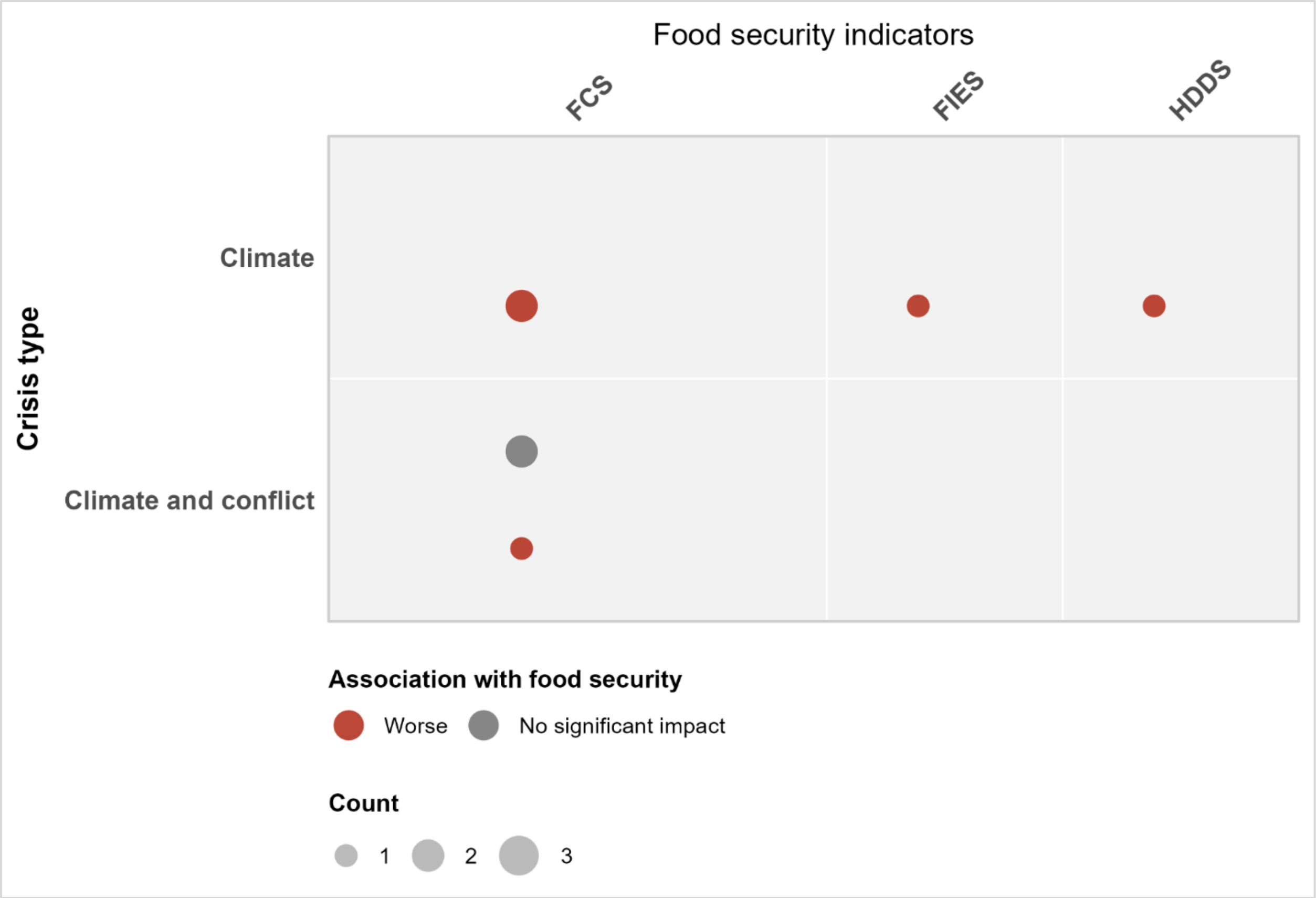

**Figure 5: Knowledge gap map generated from structured data exported from three Knowledge Loom records.**

However, the broader application of this approach was constrained by the limited availability of reusable datasets and analytical code across the reviewed studies. As a result, we could not systematically apply infrastructure-supported integration beyond a small subset of cases. Within the terms of the proposed framework, literature-based integration could therefore be applied across the full review corpus, whereas infrastructure-supported integration remained possible only under much more restrictive conditions.

## 5 Discussion: Lessons learned for knowledge integration

Our evidence shows how the complementarity proposed in our framework operates in practice. We examine this complementarity through the framework's four key dimensions: coverage of heterogeneous evidence, structured synthesis of reported findings, verifiability of reported results, and reusability of underlying research outputs. Literature-based knowledge integration provided broad coverage and supported structured synthesis across all 37 studies in the review corpus. Infrastructure-supported integration, by contrast, was feasible only for a very small subset, because it depended on whether underlying data and analytical workflows were actually accessible, sufficiently documented, and reusable. The main reason was not the absence of infrastructure itself, but the limited availability of reusable research outputs.

A related issue concerns the gap between formal openness and actual access. In our sample, several studies stated that data were available upon request, but this rarely resulted in usable materials. Similar patterns have been reported in previous studies, which document limited success in obtaining data from authors and a decline in the availability of research materials over time (Hardwicke et al., 2018; Stodden et al., 2018; Vines et al., 2014). This matters for evidence synthesis. Systematic reviews remain the most practical way to integrate heterogeneous evidence across a broad literature, especially in interdisciplinary fields where studies differ substantially in methods, indicators, and contexts. At the same time, our findings show the limits of this mode of integration. Because literature-based synthesis relies on what publications report, it rarely allows direct engagement with the analytical process behind the findings. In this respect, our results are consistent with broader work on reproducibility and reuse, which shows that publications alone are usually insufficient for verification; access to underlying data, code, and documentation is also required (Stodden et al., 2018; Wilkinson et al., 2016). A similar pattern was reported in a recent study that attempted to replicate the findings of published systematic reviews using only their reported methods, finding that database searches, screening, and data extraction were frequently inconsistent with the original review, with incomplete reporting identified as the main driver of these discrepancies, even though the resulting summary estimates were rarely affected in a meaningful way (Hamilton et al., 2026). This illustrates the limits of replication based on reported text alone and underscores why direct access to underlying data and workflows, as explored in our case study, matters for verifying scientific findings.

Our case study also demonstrates what infrastructures can add. Where reusable outputs existed, the TIB Knowledge Loom made it possible to link findings to underlying data and workflows in a structured way and to generate a synthesis output directly from those materials. This does not show that infrastructure-supported integration is broadly feasible. Rather, it shows that such integration is possible under specific conditions.

Based on these findings, several practical conditions appear particularly important for extending knowledge integration beyond publication-level synthesis:

- **Accessible datasets:** Data should be deposited in repositories that provide direct and persistent access.
- **Executable analytical code**: Analytical scripts should be shared alongside datasets in forms that allow reproduction of results.
- **Clear documentation and metadata:** Sufficient information should be provided to explain how datasets were generated and how analyses were conducted.
- **Alignment between policy and practice:** Data availability statements should reflect actual accessibility, rather than stated availability such as "upon request."
- **Coordination and dialogue across the research ecosystem:** Progress depends on coordination among researchers, journals, funders, and research infrastructures, since the reusability of research outputs is shaped by institutional as well as technical conditions.

The findings should be interpreted with caution, given the scope of the analysis. Our study builds on a systematic review of peer-reviewed, quantitative, English-language studies. While this ensures comparability, it excludes other forms of evidence and may limit generalizability. In addition, the small number of studies for which reusable data and code could be obtained limits the empirical scope of the infrastructure-supported component, and a larger sample may yield different results. At the same time, the study has important strengths. By combining conceptual framing with an empirical case study, it moves beyond abstract discussion and demonstrates concretely how literature-based and infrastructure-supported approaches can complement one another. This provides a strong and grounded basis for understanding not only where current limits remain, but also how knowledge integration can be extended in practice.

## 6 Conclusion

In this paper, we developed a conceptual framework linking literature-based synthesis with infrastructure-supported knowledge integration and used a systematic review as a case study to examine the conditions under which knowledge integration can be extended beyond reported findings. The framework shows that literature-based integration primarily supports coverage and structured synthesis, whereas infrastructure-supported integration is critical for verifiability and reusability, where underlying research outputs are accessible and usable.

From this analysis, we draw three main conclusions for knowledge integration across studies. First, the scope of integration is shaped by the condition of research outputs. In our case, integration across studies remained largely limited to reported findings, because reusable data and code were available for only a small subset of studies, even where availability was formally indicated. This indicates that integrating knowledge beyond reported findings depends not only on the existence of relevant evidence but also on whether research outputs are accessible and reusable in practice.

Second, literature-based integration remains essential because it provides the most broadly applicable basis for achieving coverage and structured synthesis across heterogeneous evidence. It enables structured comparison across studies and helps identify patterns and gaps in a fragmented literature. In our study, this was reflected in the construction of a knowledge gap map based on manually extracted information from published articles. At the same time, this mode of integration remains tied to textual reporting and therefore allows only limited engagement with the analytical processes behind published findings.

Third, research infrastructures can extend knowledge integration where reusable outputs exist. Our case study shows that synthesis outputs can, in principle, be generated directly from structured data and code rather than reconstructed from publications. However, this was feasible for only a small number of studies. The main constraint was therefore not the absence of infrastructure itself, but the limited availability of accessible, executable, and sufficiently documented research outputs.

Taken together, our findings provide empirical support for the framework's proposition that literature-based knowledge integration and infrastructure-supported knowledge integration are complementary rather than competing approaches. One supports coverage and structured synthesis across a heterogeneous literature, whereas the other supports verifiability and reusability where reusable research outputs exist. In this sense, our empirical findings clarify both the current limits of knowledge integration and the conditions under which we can extend it beyond reported findings. We therefore conclude that advancing knowledge integration depends not only on infrastructures, but also on whether data and code are shared in forms that are accessible, executable, and sufficiently documented for reuse.

## Declarations

### Ethics approval and consent to participate

Not applicable

### Consent for publication

Not applicable

### Availability of data and materials

The main scientific findings presented in this article are published together with related data and code in reproducible form in the TIB Knowledge Loom (Stocker et al., 2025) by Awoke & Brück (2026b).

## Competing interests

The authors have no competing interests to disclose.

## Funding

This research was funded by the Leibniz Lab "Systemic Sustainability" (GA LL-2024-SYSTAIN) and by the Deutsche Forschungsgemeinschaft (DFG, German Research Foundation) – project number BR 3614/4 − 1. The funders had no role in study design, data collection, data analysis, data interpretation, writing, or the decision to submit the manuscript.

## Authors' contributions

Conceptualization: MDA, HG, LS, MS, TB; Methodology: MDA, HG, LS, MS, TB; Writing-Original draft: MDA; Writing-Reviewing and editing: MDA, HG, LS, MS, TB, Supervision and funding acquisition: MS & TB

## Acknowledgements

Not applicable